\documentclass[aps,prb,twocolumn,superscriptaddress,showpacs]{revtex4-1}

\usepackage{graphicx}
\usepackage{calc}
\usepackage{bm}
\usepackage{color}
\usepackage{textcomp}

\begin{document}

\title{Band gap reconstruction at the interface between black phosphorus and a gold electrode.}

\author{N.N. Orlova}
\author{N.S. Ryshkov}
\author{A.A. Zagitova}
\author{V.I. Kulakov}
\author{A.V. Timonina}
\author{D.N. Borisenko}
\author{N.N. Kolesnikov}
\author{E.V. Deviatov}

\affiliation{Institute of Solide State Physics Russian Academy of Sciences, 142432, Chernogolovka, Moscow District, Russia}

\date{\today}

\begin{abstract}
We  experimentally investigate charge transport through the interface between a gold electrode and a black phosphorus single crystal. The experimental $dI/dV(V)$ curves are characterized by well developed zero-bias conductance peak and two strongly different branches. We  find that two branches of asymmetric $dI/dV(V)$ curves  correspond to different band gap limits, which is consistent with the theoretically predicted band gap reconstruction at the surface of black phosphorus under electric field. This conclusion is confirmed by experimental comparison  with the symmetric curves for  narrow-gap (WTe$_2$) and wide-gap (GaSe) metal-semiconductor structures.  In addition, we demonstrate p-type dopants redistribution at high bias voltages of different sign, which opens a way to use the interface structures with black phosphorus in resistive memory applications.
\end{abstract}


\maketitle

\section{Introduction}

Recent interest~\cite{BPreview,BPreview1} to black phosphorus (BP) is mostly due to  general attention to two-dimensional materials. The bulk BP is a layered semiconductor, so few- and monolayer samples (phosphorene) can be exfoliated by standard techniques.  Because of the band gap values,  BP covers the energy interval between graphene and transition metal dechalcogenides.  The bulk material is characterized by 0.3~eV band gap~\cite{BPbulk,BPbulk2}, which can be significantly increased (up to 2~eV) in monolayer structures~\cite{BP1,BP2,BP3}.  BP is expected to have potential applications in the fields of flexible electronics~\cite{BPr1,BPr2}, gas sensing~\cite{BPr3}, photodetection~\cite{BPr4,BPr5}, catalysis~\cite{BPr6,BPr7}, imaging~\cite{BPr8}, plasmonics~\cite{BPr9}.

A unique property of phosphorene is high tunability of the band gap by strain~\cite{strain1,strain2} and perpendicular electric field~\cite{field1,field2}. In the latter case, the band gap increases due to  the Stark effect for positive fields, whereas in the negative fields the band gap disappears~\cite{2Dfield}. This band gap reconstruction leads to the semiconductor-to-semimetal and semiconductor-to-metal phase transitions, the critical field values can be estimated~\cite{Ecrit2D} as 0.15-0.17 V/\AA. 

Doping  can significantly affect phosphorene properties~\cite{BPr13,BPr14,BPr15,BPr16,BPr17}. On the other hand, BP monolayers are known to be unstable under ambient conditions~\cite{ambient,doping}. The exfoliation methods exhibit low efficiency, especially for heterostructures with boron nitride~\cite{hBn}, so phosphorene is obviously not suitable for industrial use. For this reason, there is significant interest to bulk and surface properties of three-dimensional BP crystals.  Band gap reconstruction has been also predicted theoretically for bulk BP, the gap disappears at  $\approx$2\% compressive strain values~\cite{strainBulk} or  $\approx$0.34~V/\AA  ~perpendicular electric fields~\cite{katsnelson1,katsnelson2}. This electric field is too strong for experimental realization in the bulk of a thick sample, however, it can be obtained at the BP crystal surface. For example, deposition of potassium atoms on the surface of single-crystalline black phosphorus modulates~\cite{field_bulk_exp1,field_bulk_exp2} the band gap in the wide range of 0.0-0.6 eV.  The mechanism of this band reconstruction has been identified as the giant Stark effect~\cite{2Dfield} due to strong vertical electric fields induced by potassium atoms~\cite{field_bulk_exp1,field_bulk_exp2}. Even band inversion can be achieved for surface deposition of K, Rb or Cs atoms~\cite{inversion1,inversion2}.

Significant electric fields are also achievable at the interface between a BP crystal and a metallic electrode: all the applied voltage imbalance should drop at the interface due to the electric field screening by bulk BP carriers~\cite{screening}. The metallic electrode can also protect the BP surface from oxidation, which significantly simplifies the requirements to the experimental setup. Usually, investigations of the semiconductor/metal interface provide significant information about semiconductor surface spectrum~\cite{schottky,ingasb}

On the other hand, the bulk p-type BP conductance is known~\cite{bulk} to be caused by vacancies in the BP lattice~\cite{vacancies1,vacancies2}. Because of low BP melting temperature, one can expect significant vacancy migration at high current densities. The corresponding change of the interface resistance can be used for resistive memory development, which is extremely important nowadays~\cite{mem1,mem2}.

Here, we experimentally investigate charge transport through the interface between a gold electrode and a black phosphorus single crystal. The experimental $dI/dV(V)$ curves are characterized by well developed zero-bias conductance peak and two strongly different branches. We  find that two branches of asymmetric $dI/dV(V)$ curves  correspond to different band gap limits, which is consistent with the theoretically predicted band gap reconstruction at the surface of black phosphorus under electric field. This conclusion is confirmed by experimental comparison  with the symmetric curves for  narrow-gap (WTe$_2$) and wide-gap (GaSe) metal-semiconductor structures.  In addition, we demonstrate p-type dopants redistribution at high bias voltages of different sign, which opens a way to use the interface structures with black phosphorus in resistive memory applications.

\section{Samples and techniques}

\begin{figure}[t]
\center{\includegraphics[width=\columnwidth]{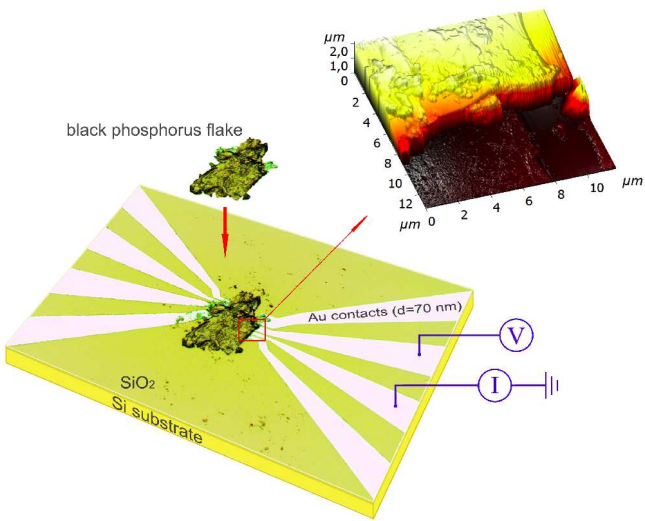}}
\caption{Optical image of a sample. A black phosphorus flake is transferred on a top of Au leads pattern. The leads are of 10,10,20,40 $\mu$m widths, respectively, they are separated by 5~$\mu$m intervals. AFM image of the flake edge is demonstrated in the inset. For correct measurements of low-conductive samples,  voltage  $V$ is directly applied between two neighbor contacts. The applied voltage is modulated by a small ac component, a lock-in measures the ac component of the current $I$, which is proportional to the differential conductivity $dI/dV$. Two Au-BP interfaces are necessary different due to the non-perfect alignment in flake deposition and different widths of  Au leads, so the experimental $dI/dV(V)$ curves mostly reflect a single, the most resistive gold-semiconductor interface~\cite{schottky}.
}
\label{sample}
\end{figure}

The BP single crystals were synthesized by the gas-transport reaction method from  red phosphorus using AuSn and SnI$_4$ as mineralization additives in a short way transport reaction ~\cite{synthesis}. The X-ray diffraction analysis confirms an orthorhombic crystalline structure of synthesized crystals (\emph{Cmca} space group, a=4.37, b=3.31, c=10.47 ~\AA). The layered structure of BP crystals makes them suitable for mechanical exfoliation, e.g. with a scotch-tape technique. To avoid sample degradation, rather thick flakes (about 100~$\mu$m size and 1$\mu$m thick) should be chosen, see Fig.~\ref{sample}. 

We select the flakes with suitable area and a clean surface, which was verified by an optical microscope. Then, a small flake is transferred to the Au leads pattern, as depicted in Fig.~\ref{sample}. The leads pattern is defined on the insulating SiO$_2$ substrate by a lift-off technique after thermal evaporation of 70~nm Au. The leads are of 10,10,20,40 $\mu$m widths, respectively, they are separated by 5~$\mu$m intervals. The selected black phosphorus  flake is pressed to the leads pattern with another oxidized silicon substrate. A weak pressure is applied with a special metallic frame, which keeps the substrates strictly parallel. This procedure provides electrically stable contacts~\cite{nbwte,cdas,magnon} with high quality interfaces~\cite{inwte,cosnsjc}. For example, it was demonstrated for Cd$_3$As$_2$ crystals, that Au-Cd$_3$As$_2$ interface quality is even higher for this technique  than for contacts directly evaporated on a polished crystal surface~\cite{cdas}.

The same technique is applied to obtain reference samples, where GaSe or WTe$_2$ thick flakes are situated on similar Au leads. GaSe is a wide-gap (2.01~eV) layered p-type semiconductor, which allows direct comparison with the black phosphorus structures. In contrast, WTe$_2$ is a topological semimetal with low bulk resistance due to the band touching in several Weyl points. Synthesis of WTe$_2$ and GaSe single crystals has been described elsewhere~\cite{nbwte,inwte,gase,gase1}.

For correct measurement of low-conductive samples one have to directly define voltage bias $V$ and measure the arising current $I$ in the circuit. Thus, it is impossible to avoid a two-point connection scheme: one of the contacts in Fig.~\ref{sample} is grounded, the applied voltage varies within $\pm$10~V range at the neighbor contact, which corresponds to the maximum current of about 0.1~mA through the sample ($\approx$100~kOhm resistance). To obtain differential conductivity curves $dI/dV(V)$, the applied voltage $V$ is additionally modulated by a small (5-10~mV) ac component at a frequency of 1100 Hz. The ac current component is measured by lock-in, being proportional to differential conductivity $dI/dV$ at a given bias voltage $V$. We verify that the obtained $dI/dV$ value is independent of the modulation frequency in the range 1~kHz--10~kHz, which is determined by the applied filters.  All the measurements are carried out at room temperature, because of known~\cite{BPbulk,BPbulk2,field_bulk_exp1,field_bulk_exp2} 0.3~eV band gap for bulk black phosphorus.

\section{Experimental results}

\begin{figure}[t]
\center{\includegraphics[width=\columnwidth]{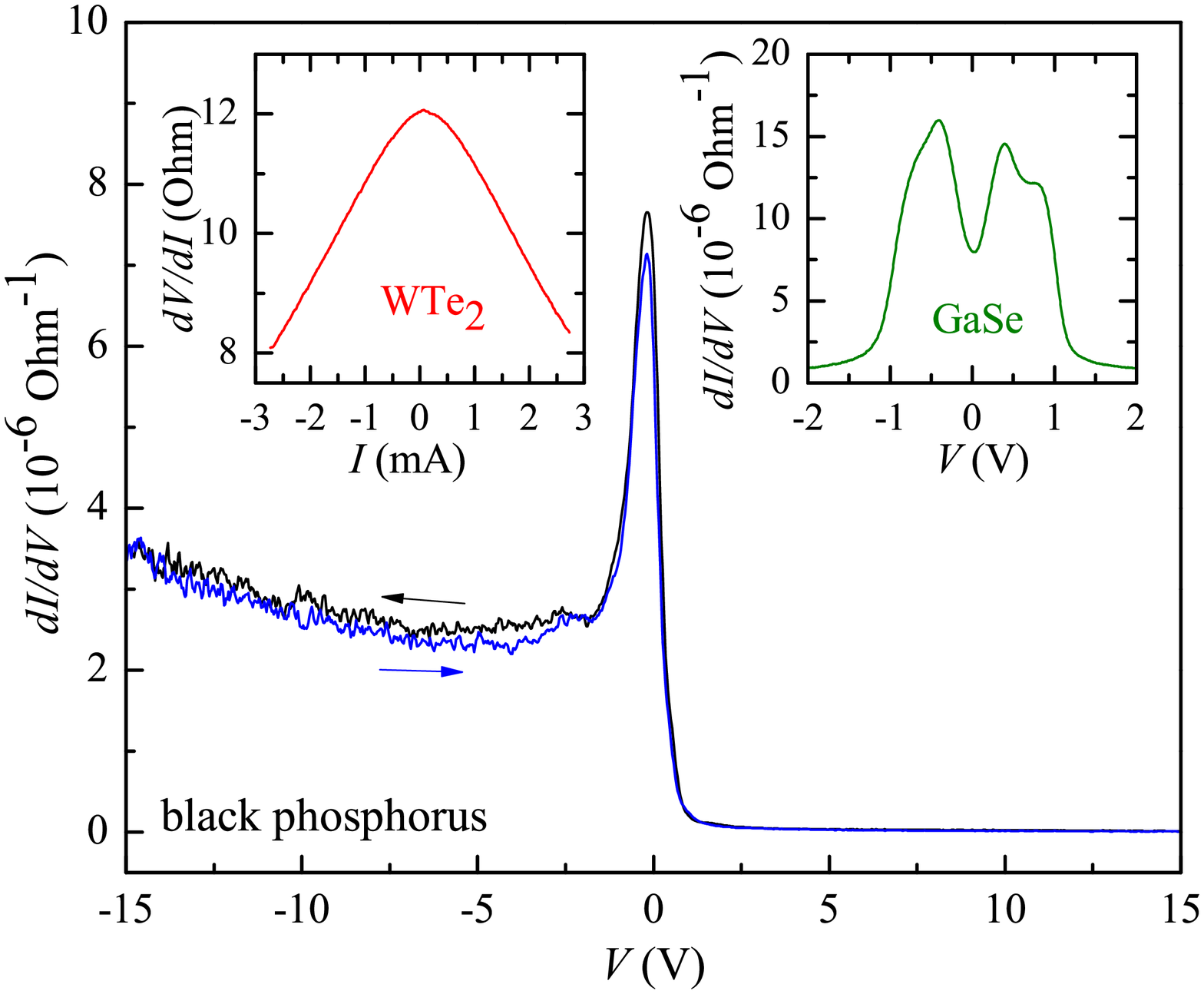}}
\caption{Examples of experimental $dI/dV(V)$ curves for the Au-BP sample. The curves are strongly asymmetric, they are characterized by well developed zero-bias conductance peak and two strongly different branches. There is  visible  hysteresis for the negative (left) $dI/dV(V)$ branch  for two different voltage sweep directions, the hysteresis is accompanied by two distinct zero-bias resistance values.  The observed $dI/dV(V)$ asymmetry is very unusual and specific for black phosphorus. The insets demonstrate experimental curves for the reference Au-GaSe and Au-WTe$_2$ samples, which are wide-gap and zero-gap semiconductors. The curves are symmetric due to the electron-hole symmetry, they reflect the band gap at the interface. }
\label{IV}
\end{figure}

Fig.~\ref{IV} shows examples of experimental $dI/dV(V)$ curves. The curves are strongly asymmetric, they are characterized by well developed zero-bias conductance peak and two strongly different branches. The zero-bias conductance value corresponds to about 150~kOhm sample reesistance. The right (positive) $dI/dV(V)$ branch demonstrates zero differential conductance for biases above 1~V. In contrast, for negative voltages, differential conductance drops in 2.5 times only, so it is obviously finite at any bias values. The left (negative) $dI/dV(V)$ branch is also  non-linear: differential conductance is gradually increasing for higher negative biases.  

We wish to note, that for resistive samples in a two-point connection scheme, the sample resistance is provided by two semiconductor-gold interfaces and a part of the bulk semiconductor. Since there is no intrinsic asymmetry for the bulk BP material, the observed $dI/dV(V)$ asymmetry  can not be connected with the bulk effects. The  $dI/dV(V)$ asymmetry would also impossible for two strictly equivalent semiconductor-gold interfaces. On the other hand, two Au-BP interfaces are necessary different due to the non-perfect alignment in flake deposition and different widths of  Au leads, see  Fig.~\ref{sample}. Both interfaces are still conducting at zero bias, the applied voltage $V$ drops mostly on the most resistive one.  This non-equivalence is even more important at finite biases in the case of strongly non-linear $dI/dV(V)$ characteristics, like it is presented in Fig.~\ref{IV}. 

Thus, the experimental $dI/dV(V)$ curves mostly reflect a single, the most resistive gold-semiconductor interface, so the two-point technique is quite appropriate  in this case. The applied  technique is similar to one in Ref.~\onlinecite{schottky}, where it was employed to Schottky and Ohmic contacts investigation in the two-point circuit with contacts of strongly different areas. It is obvious, that for the two-point connection scheme, the sign of the applied voltage, and, therefore, the sign of $dI/dV(V)$ curve asymmetry,  is defined by the choice of the connection leads.

\begin{figure}[t]
\center{\includegraphics[width=\columnwidth]{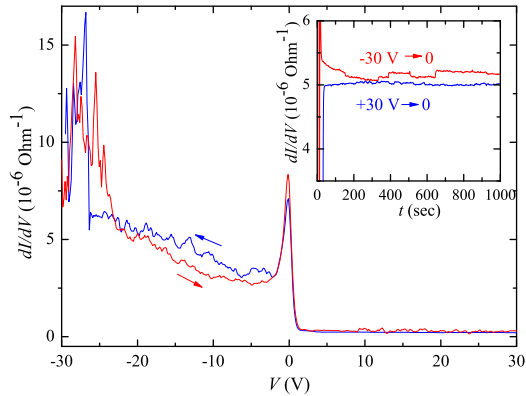}}
\caption{Hysteresis for a wider voltage sweep range, which leads to the larger  $\Delta dI/dV(V=0)\approx 1\cdot$10$^{-6} \Omega^{-1}$. The inset demonstrates the relaxation curves $dI/dV(t)$ at  $V=0$ for two high dwelling biases ($\pm$30~V). The final stable $dI/dV(V=0)$ value obviously depends on the sign of dwelling bias,  while the initial $\Delta dI/dV(V=0)\approx 1\cdot$10$^{-6} \Omega^{-1}$ is stabilized at $\approx 0.3\cdot$10$^{-6} \Omega^{-1}$ value.}
\label{relax}
\end{figure}

The qualitative behavior of the observed $dI/dV(V)$ asymmetry is very unusual. For a standard Ohmic contact, both $dI/dV(V)$ branches are characterized by finite differential conductance. For a  Schottky contact, it should be no current for one of the $I-V$ branches, see, e.g., Ref.~\onlinecite{schottky}. In contrast, zero differential conductance beyond the symmetric zero-bias conductance peak region in Fig.~\ref{IV} indicates current saturation at some significant value, which is not possible both for Ohmic and Schottky contacts. Also, the symmetric zero-bias peak is inconsistent with band-bending induced Schottky barrier.

Since the applied biases are well above the expected 0.3~eV band gap value in Fig.~\ref{IV}, we can expect that the experimental $dI/dV(V)$ curves  reflect the band gap behavior at the interface. For conventional semiconductors like GaSe, we could expect symmetric $dI/dV(V)$ curves, as it is demonstrated in the right inset to Fig.~\ref{IV} for a reference Au-GaSe sample. This material is characterized by 2.01~eV band gap and bulk p-type doping~\cite{gase,gase1}, which allows direct comparison with the BP behavior. The maximum number of carriers is still defined by dopants concentration (one can neglect thermally activated carriers for this band gap value), so the current saturates for Fermi level positions above the bottom of the conductance band. This is reflected by zeroing the differential conductance at finite biases, which we indeed observe for GaSe in the inset to Fig.~\ref{IV}. The proposed explanation should be symmetric in respect to the bias sign due to the obvious electron-hole symmetry. In the right inset to Fig.~\ref{IV}, $dI/dV$ is finite within $\pm$1~V range for our  GaSe samples, which well corresponds to the known~\cite{gase,gase1} 2.01~eV band gap.  This experimental observation supports the conclusion that $dI/dV(V)$ curve should  reflect the gap value for a single gold-semiconductor interface.

Due to the electron-hole symmetry, one should also expect  symmetric curves for zero-gap materials, like Weyl semimetal WTe$_2$. In the case of a reference Au-WTe$_2$  sample, the current is applied and the voltage is measured~\cite{nbwte} to obtain $dV/dI(I)$, because of low (much below 100~Ohm) sample resistance. The experimental $dV/dI(I)$ curve is indeed symmetric  in the left inset to Fig.~\ref{IV} for Au-WTe$_2$. The curve is  monotonous, since there is no gap in the WTe$_2$ spectrum. The resistance is diminishing for higher biases, like we obtain for the left  $dI/dV(V)$ branch in the main field of Fig.~\ref{IV} for BP. It seems that two branches of the experimental $dI/dV(V)$ curves for black phosphorus resemble two different limits (finite and zero) of the band gap at the interface.

\begin{figure}[t]
\center{\includegraphics[width=\columnwidth]{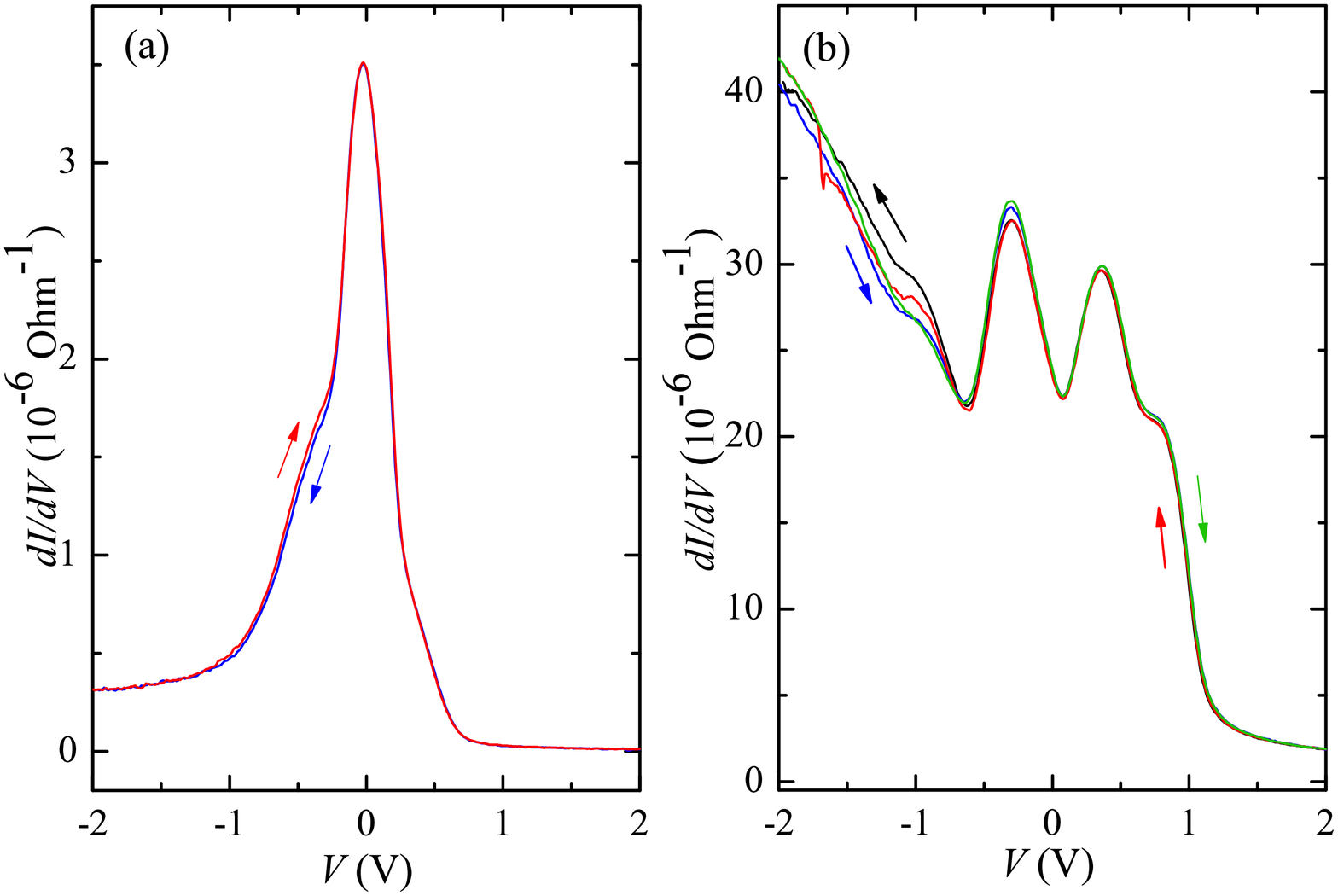}}
\caption{  Maximum device-to-device fluctuations in our experiment.  (a) For the most resistive sample, $dI/dV(V)$ curves are characterized by single zero-bias peak, the branch asymmetry is smaller than for the sample in Figs.~\ref{IV},~\ref{relax}. (b) For the most conducting sample, the symmetric zero-bias conductance peak has a local minimum, the branch asymmetry is much higher. In both the cases, $dI/dV(V)$ curves are qualitatively similar to ones in Fig.~\ref{IV}, they also demonstrate the same $\approx$1~V threshold voltage for the low-conductive branch. The curves are shown for two sweep directions.  Because of the narrow voltage range, the hysteresis is negligible, the curves are well reproducible in different scans, as it is demonstrated by two up and two down scans in (b).  
}
\label{IVdiff}
\end{figure}

Fig.~\ref{IV} also presents experimental $dI/dV(V)$ curves for two different voltage sweep directions. There is  visible  hysteresis for the negative (left) $dI/dV(V)$ branch in Fig.~\ref{IV}. This hysteresis is accompanied by two distinct zero-bias resistance values, the difference $\Delta dI/dV(V=0)$ is about 0.5$\cdot$10$^{-6} \Omega^{-1}$. For the positive branch, zero differential conductance does not depend on the sweep direction. The curves are well reproducible in different voltage scans, both the hysteresis magnitude and the difference $\Delta dI/dV(V=0)$ slightly depend on the voltage sweep rate.

We observe, that $\Delta dI/dV(V=0)$ is increased for a wider voltage sweep range, see Fig.~\ref{relax}. There are also strong instabilities at high negative biases (below -20~V), which are not present for positive ones. We should conclude, that the interface properties are modified by high current values, which are easily achievable for the finite differential conductance branch. It leads to the larger  $\Delta dI/dV(V=0)\approx 1\cdot$10$^{-6} \Omega^{-1}$ in Fig.~\ref{relax}. 

This conclusion is supported by the relaxation curves $dI/dV(t)$ at zero bias $V=0$ in the inset to Fig.~\ref{relax}. To obtain the $dI/dV(t)$ curves, the sample is kept for about 1 minute  at high dwelling bias ($\pm$30~V). After that,   the initial bias is set to zero and  the differential conductivity $dI/dV(V=0)$ is traced in dependence of time. We do not observe visible relaxation for the positive dwelling bias $+30$~V, while some $dI/dV(t)$ dependence can be seen for the negative $-30$~V one. The final stable $dI/dV(V=0)$ value obviously depends on a sign of dwelling bias in Fig.~\ref{relax}, while the initial $\Delta dI/dV(V=0)\approx 1\cdot$10$^{-6} \Omega^{-1}$ is stabilized at the smaller $\approx 0.3\cdot$10$^{-6} \Omega^{-1}$ value.

\begin{figure}[t]
\center{\includegraphics[width=\columnwidth]{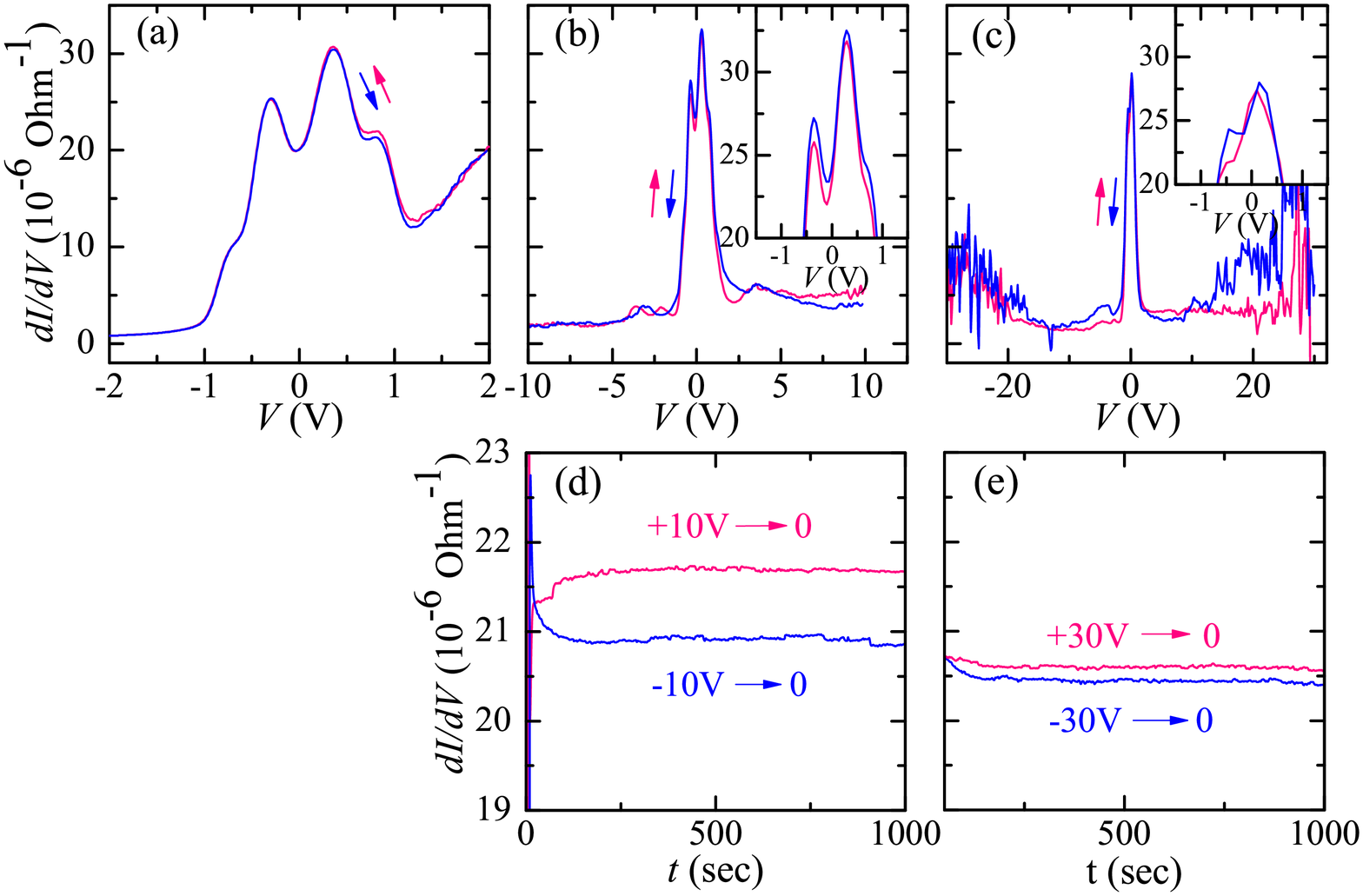}}
\caption{ Hysteresis effects and sample modification by high biases for a different (the forth one) sample. (a) For the narrow voltage sweep range, $dI/dV(V)$ curves are qualitatively similar to ones in Fig.~\ref{IV},~\ref{IVdiff}, they also demonstrate the same 1~V threshold voltage for the low-conductive branch.  The zero-bias $dI/dV(V=0)$ value is independent of the sweep direction. For the two-point connection scheme, the sign of the applied voltage  is defined by the choice of the connection leads. (b) Considerable hysteresis appears for the  $\pm$10~V voltage range, the zero-bias conductance $dI/dV(V=0)$ demonstrates two distinct values for two sweep directions, see the inset. (c) $dI/dV$ instabilities appears for the widest bias range, $dI/dV(V)$ curves are even more symmetric. The difference of zero-bias $dI/dV(V=0)$ values is diminished, see inset. (d-e) Relaxation curves for different dwelling voltages, $\pm$10~V and $\pm$30~V, respectively, which allows to estimate the stability of $dI/dV(V=0)$ and final $\Delta dI/dV(V=0)$ values. 
}
\label{IVsecond}
\end{figure}

Figs.~\ref{IVdiff},~\ref{IVsecond} (a)  allow to estimate maximum device-to-device fluctuations in our experiment.   For the most resistive sample ($dI/dV(V=0)\approx 3.5\cdot$10$^{-6} \Omega^{-1}$), $dI/dV(V)$ curves are characterized by single zero-bias peak, see Fig.~\ref{IVdiff} (a), the branch asymmetry is smaller than for the sample in Fig.~\ref{IV}. For the most conducting sample in Fig.~\ref{IVdiff} (b), $dI/dV(V=0)\approx 30\cdot$10$^{-6} \Omega^{-1}$,  the symmetric zero-bias conductance peak has an additional local minimum. The $dI/dV(V)$ curve still consists from the symmetric zero-bias conductance region and two different branches, one of them is of constant, nearly zero conductance. The other $dI/dV(V)$ branch demonstrates significant differential conductivity, which is increasing at high biases. Thus, we not only observe qualitatively similar $dI/dV(V)$ curves for samples of different resistance, but also demonstrate the same $\approx$1~V threshold voltage for low-conductive branch in Figs.~\ref{IV},~\ref{IVdiff} and \ref{IVsecond} (a). 

Fig.~\ref{IVsecond} demonstrates hysteresis effects and sample modification by high biases. For the narrow voltage sweep range in Fig.~\ref{IVsecond} (a), $dI/dV(V)$ curves are similar to ones for the sample from Fig.~\ref{IVdiff} (b). There is only small hysteresis for positive, high-conductive, $dI/dV(V)$ branch, the zero-bias $dI/dV(V=0)\approx 30\cdot$10$^{-6} \Omega^{-1}$ value is independent of the sweep direction. Similarly to the first sample, considerable hysteresis appears for $\pm$10~V voltage scans,  as depicted in Fig.~\ref{IVsecond} (b). The zero-bias $dI/dV(V=0)$ value is of  two distinct values for two sweep directions, see the inset to Fig.~\ref{IVsecond} (b). The stability of these values and final $\Delta dI/dV(V=0)$ can be estimated from the relaxation curves in Fig.~\ref{IVsecond} (d). 

The overall $dI/dV(V)$ asymmetry is diminished for the  $\pm$10~V voltage range in Fig.~\ref{IVsecond} (b), it more corresponds to the asymmetry of experimental curves for the first sample, cp.  Figs.~\ref{IV} and~\ref{IVsecond} (b). For higher biases, also in a good correspondence with the first sample behavior, $dI/dV$ instabilities appear, which can be seen in Fig.~\ref{IVsecond} (c). The $dI/dV(V)$ curves are more symmetric in this voltage range, the difference of zero-bias $dI/dV(V=0)$ values is diminished, see the inset to Fig.~\ref{IVsecond} (c) and the relaxation curves in Fig.~\ref{IVsecond} (e).

\section{Discussion}

As a result, we obtain strongly asymmetric $dI/dV(V)$ curves with a developed zero-bias conductance peak for transport through Au-BP interface.  In addition, for high dwelling voltages $V$,  we observe  two stable zero-bias $dI/dV(V=0)$ values, which  are determined by a sign of the initially applied bias. 

The second effect can be easily explained by vacancy diffusion at high currents. Indeed, black phosphorus is a p-type semiconductor, due to vacancy centers in the crystal lattice~\cite{bulk}. Black phosphorous decomposition is known  to occur at sufficiently low temperatures, 400 -- 500$^\circ$~C, it  continues until only a thin, amorphous red phosphorous remains~\cite{decomp}. As usual, vacancy diffusion is significant near the melting point, so, one can expect  vacancy redistribution  at the interface in the electric field of flowing current because of Joule heating. The heating effects are significant for the conductive $dI/dV(V)$ branch, while current saturates for the other one. This mechanism  is confirmed by $dI/dV(V)$ modification in  Fig.~\ref{IVsecond} (c), where the symmetric $dI/dV(V)$ curve is restored.

The situation is more complicated with the  observed $dI/dV(V)$ asymmetry for black phosphorous samples. First of all, experimental $dI/dV(V)$ curves can not be explained by Schottky effect in Figs.~\ref{IV} and~\ref{IVsecond}. Both for Au-Bp and Au-GaSe samples, we observe symmetric zero-bias conductance peak, which is accompanied by current saturation above some threshold voltage. This is just opposite to that one can expect both for Ohmic and Schottky contacts~\cite{schottky}, as well as for the potential barrier at the interface.  
 
In the case of a wide bias range, $dI/dV(V)$ curve should reflect the band gap as a region of non-zero differential conductance, as it is demonstrated for p-type GaSe sample in the right inset to Fig.~\ref{IV}. We wish to note, that GaSe sample is of similar zero-bias conductance value and is measured in the same two-point technique. In contrast, we observe two strongly different branches in Figs.~\ref{IV} and~\ref{IVsecond} (a). For both Au-BP samples, $dI/dV$ approaches zero value above 1~V, which obviously exceeds the phosphorus bulk band gap 0.3~eV. For the opposite branch, differential conductance is rising with voltage imbalance, as it is known for zero-gap semiconductors, see the left inset to Fig.~\ref{IV}. 

Thus, from the comparison with Au-WTe$_2$ and Au-GaSe semiconductor structures, we can conclude that two branches of the experimental $dI/dV(V)$ curves for black phosphorus reflect two different limits of the band gap,  the  zero-gap and wide-gap ones, respectively. This conclusion  well corresponds to the predicted~\cite{katsnelson1,katsnelson2} energy gap reconstruction in black phosphorus due to strong vertical electric fields and to the observed~\cite{field_bulk_exp1,field_bulk_exp2} band gap modulation  in the wide range of 0.0-0.6 eV. The band gap value is expected~\cite{2Dfield} to be increased for one field direction to the layers, or it even disappears for the other one due to  the Stark effect.  We should consider 1~mV threshold voltage in Figs.~\ref{IV} and~\ref{IVsecond} (a) as a result of band gap increase in three times from the initial 0.3~eV bulk value. On the other hand, the  band gap is predicted to 
disappear at $\approx$0.34~V/\AA ~field strength~\cite{katsnelson1,katsnelson2}, which  can be easily obtained at the interface. Indeed, experimental $dI/dV(V)$ curves confirm finite  bulk  conductance for BP. Thus, an external electric field is screened at distances, which are of the order of the interlayer ones~\cite{screening,length}, so, the applied bias produces a voltage drop of the comparable value directly at the Au-BP interface. 

Similar voltage drop is also demonstrated in experiments as band bending near the surface~\cite{screening,adatoms}.   A sufficiently strong band bending can even lead to confinement of the conduction band states near the interface~\cite{field_bulk_exp1,inversion2,bending1,bending2}. Band bending could distort our estimation of the gap value, but it can not revise the conclusion on the band gap reconstruction: since bend bending is the result of intrinsic electric field at the interface~\cite{screening,adatoms}, it can only produce $dI/dV(V)$ curves shift in $V$-direction due to the external field compensation.  In contrast, the conductance peak is well-centered at zero bias both for Au-BP and Au-GaSe samples in Figs.~\ref{IV},~\ref{IVdiff}, and~\ref{IVsecond},  so the band bending induced shift is not noticeable for this wide voltage range. On the other hand, it is well known, that confinement of the conduction band states is important for transport along the surface, while band reconstruction is for vertical (normal to the interface) transport~\cite{2DEGtransport}, which we investigate here. 

Modification of $dI/dV(V)$ curves by high currents in Fig.~\ref{IVsecond} is an additional argument: the zero-bias peak survives even after treatment of the sample by high currents, which, due to vacancy migration, should diminish electric field (and band bending) at the interface. Thus, the zero-bias conductance is indeed defined by BP doping, while the initial $dI/dV(V)$ asymmetry is connected with electric field.

\section{Conclusion}

In conclusion, we experimentally investigate charge transport through the interface between a gold electrode and a black phosphorus single crystal. The experimental $dI/dV(V)$ curves are characterized by well developed zero-bias conductance peak and two strongly different branches. We  find that two branches of asymmetric $dI/dV(V)$ curves  correspond to different band gap limits, which is consistent with the theoretically predicted band gap reconstruction at the surface of black phosphorus under electric field. This conclusion is confirmed by experimental comparison  with the symmetric curves for  narrow-gap (WTe$_2$) and wide-gap (GaSe) metal-semiconductor structures.  In addition, we observe  two distinct zero-bias $dI/dV(V=0)$ values if the black phosphorus sample is initially subjected to high bias voltages of different sign. The latter effect can be explained by p-type dopants redistribution, it opens a way to use the interface structures with black phosphorus in resistive memory applications.

\section{Acknowledgement}
The authors are grateful to V.T. Dolgopolov and O.O. Shvetsov for fruitful discussions, S.V. Chekmazov, A.M. Ionov  and S.S. Khasanov for STM, XPS and X-ray black phosphorus characterization. We gratefully acknowledge financial support partially by the RFBR  (project No.~19-29-03021),  and RF State task.

\end{document}